\documentclass[aps,prd,onecolumn,groupedaddress,showpacs,nofootinbib,amssymb]{revtex4-2}
\usepackage[dvips]{graphicx}
\usepackage{amssymb}
\usepackage{amsmath}
\usepackage{graphicx,,color}
\usepackage{amsfonts}
\usepackage{bm}
\usepackage{cancel}
\usepackage{comment}

\newcommand\be{\begin{equation}}
\newcommand\ee{\end{equation}}

\allowdisplaybreaks[4]

\begin{document}

\title{Effects of a Pre-inflationary de Sitter Bounce on the Primordial Gravitational Waves in $f(R)$ Gravity Theories}
\author{V.K. Oikonomou,$^{1}$}
\email{v.k.oikonomou1979@gmail.com,voikonomou@auth.gr}
\affiliation{$^{1)}$Department of Physics, Aristotle University of
Thessaloniki, Thessaloniki 54124, Greece}


\tolerance=5000

\begin{abstract}
In this work we examine the effects of a pre-inflationary de
Sitter bounce on the energy spectrum of the primordial
gravitational waves. Specifically we assume that the Universe is
described by several evolution patches, starting with a de Sitter
pre-inflationary bounce which is followed by an quasi-de Sitter
slow-roll inflationary era, followed by a constant equation of
state parameter abnormal reheating era, which is followed by the
radiation and matter domination eras and the late-time
acceleration eras. The bounce and the inflationary era can be
realized by vacuum $f(R)$ gravity and the abnormal reheating and
the late-time acceleration eras by the synergy of $f(R)$ gravity
and the prefect matter fluids present. Using well-known
reconstruction techniques we find which $f(R)$ gravity can realize
each evolution patch, except from the matter and radiation
domination eras which are realized by the corresponding matter
fluids. Accordingly, we calculate the damping factor of the
primordial de Sitter bounce, and as we show, the signal can be
detected by only one gravitational wave future experiment, in
contrast to the case in which the bounce is absent. We discuss in
detail the consequences of our results and the future
perspectives.
\end{abstract}


\maketitle

\section{Introduction}

The two acceleration eras of our Universe, including the reheating
era, are undoubtedly the most mysterious eras of all the
evolutions eras we assume that our Universe experienced. The
late-time acceleration era though is confirmed and the difficulty
lies to pinpointing the physical theory and  mechanism which
controls it. However, the other two eras are speculated to have
occurred in the primordial epoch of the Universe, and to date no
firm evidence is provided that these eras have actually occurred.
With regard to the inflationary era
\cite{inflation1,inflation2,inflation3,inflation4}, the occurrence
of this era could be verified by the direct detection of the
B-modes in the Cosmic Microwave Background (CMB) temperature
fluctuations. This for example can be verified in the stage 4 CMB
experiments \cite{CMB-S4:2016ple,SimonsObservatory:2019qwx} in
some years from now, or alternatively, the primordial stochastic
tensor modes can be directly detected in future gravitational
waves experiments
\cite{Hild:2010id,Baker:2019nia,Smith:2019wny,Crowder:2005nr,Smith:2016jqs,Seto:2001qf,Kawamura:2020pcg,Bull:2018lat},
see also Ref. \cite{LISACosmologyWorkingGroup:2022jok} for some up
to date information on cosmological studies of the LISA mission.

Now the plot may thicken with the existence or not of the
inflationary era. The results of the future CMB and gravitational
waves experiments will play a crucial role, in both cases of
detection or not of a signal. In the unlikely event of
non-observation of a signal, the scientists will be confronted
with the difficult task to explain why no signal is detected. Is
this non-detection because inflation did not occur, or simply
because inflation is described by a theory which yields a negative
tensor spectral index and a non-detectable by the current
experiments signal, while it also yields a standard General
Relativistic (GR) reheating era? On the antipode of this, there
lies the detection of a signal. Many questions can be asked then,
how strong is the signal, is it detectable by several experiments
in various frequency ranges, or by some of the experiments? With
regard to how strong a signal can be, this is very important. The
observation of a signal in some but not all the detectors may
signify some physical process which causes damping of the signal,
such as supersymmetry breaking after or during reheating. This
will also determine the era for which the physics change occurred
in the Universe. However, with regard to how strong the signal is,
many things can be said, and many questions can be asked. The
answers to these questions may vary and strongly depend on the
final form of the signal. Thus the strength and form of the signal
may determine whether this signal is obtained by a theory with
positive tensor spectral index, or by a standard inflationary
theory with an abnormal reheating era. In the literature, many
aspects on primordial gravitational waves are studied
\cite{Kamionkowski:2015yta,Denissenya:2018mqs,Turner:1993vb,Boyle:2005se,Zhang:2005nw,Schutz:2010xm,Sathyaprakash:2009xs,Caprini:2018mtu,
Arutyunov:2016kve,Kuroyanagi:2008ye,Clarke:2020bil,Kuroyanagi:2014nba,Nakayama:2009ce,Smith:2005mm,Giovannini:2008tm,
Liu:2015psa,Zhao:2013bba,Vagnozzi:2020gtf,Watanabe:2006qe,Kamionkowski:1993fg,Giare:2020vss,Kuroyanagi:2020sfw,Zhao:2006mm,
Nishizawa:2017nef,Arai:2017hxj,Bellini:2014fua,Nunes:2018zot,DAgostino:2019hvh,Mitra:2020vzq,Kuroyanagi:2011fy,Campeti:2020xwn,
Nishizawa:2014zra,Zhao:2006eb,Cheng:2021nyo,Nishizawa:2011eq,Chongchitnan:2006pe,Lasky:2015lej,Guzzetti:2016mkm,Ben-Dayan:2019gll,
Nakayama:2008wy,Capozziello:2017vdi,Capozziello:2008fn,Capozziello:2008rq,Cai:2021uup,Cai:2018dig,Odintsov:2021kup,Benetti:2021uea,Lin:2021vwc,Zhang:2021vak,Odintsov:2021urx,Pritchard:2004qp,Zhang:2005nv,Baskaran:2006qs,Oikonomou:2022xoq,Odintsov:2022cbm,Odintsov:2022sdk,Kawai:2017kqt,Odintsov:2022hxu,Gao:2019liu,Oikonomou:2022pdf},
and with regard to the abnormal reheating perspective and effects
on the energy spectrum of the primordial gravitational waves, this
aspect has recently been studied in Refs.
\cite{Oikonomou:2022pdf,Oikonomou:2022xoq,Odintsov:2022sdk} and it
was shown that the gravitational waves energy spectrum of standard
$f(R)$ gravity inflation can be enhanced significantly by the
presence of an $f(R)$ gravity generated reheating era. This is in
contrast to a GR compatible reheating era of course and the
flatness, form and strength of the detected signal may reveal many
properties regarding the underlying theory. Regarding the
strength, this may vary and one mechanism of enhancement or
damping may be the presence of some peculiar pre-inflationary era,
see for example Ref. \cite{Odintsov:2021urx} for the effects of a
primordial bounce on the energy spectrum of the primordial
gravitational waves. Specifically, in Ref. \cite{Odintsov:2021urx}
the effects of a primordial pre-inflationary bounce on the energy
spectrum of the inflationary gravitational waves were considered.
In this work we shall also consider the effects of a
pre-inflationary de Sitter bounce on the energy spectrum of the
primordial gravitational waves, in the context of $f(R)$ gravity.
In standard string theory scenarios, pre-inflationary epochs may
actually lead to an overall amplification of the gravitational
wave energy spectrum \cite{Gasperini:2007vw}, see also Refs.
\cite{Navascues:2021mxq,Anderson:2020hgg,Li:2019ipm,Cai:2015nya,Wang:2014abh,Kitazawa:2014dya,Rinaldi:2010yp}.
This was also the case in Ref. \cite{Odintsov:2021urx}, however in
this work we shall demonstrate that it is possible a primordial
bounce to lead to an overall damping of the gravitational waves
energy spectrum, which is quite significant. In general, bouncing
cosmology
\cite{Brandenberger:2012zb,Brandenberger:2016vhg,Battefeld:2014uga,Novello:2008ra,Cai:2014bea,deHaro:2015wda}
is a possible alternative to the inflationary scenario, so in this
work we combine the presence of a pre-inflationary bounce with a
standard post-bounce slow-roll inflationary era. Our assumption is
that the Universe's dynamics is controlled by an $f(R)$ gravity
during the pre-inflationary and inflationary era, and
post-inflationary the evolution is controlled by the synergy of
$f(R)$ gravity in the presence of matter and radiation perfect
fluids. As we will show, the predicted energy density of the
primordial gravitational waves is damped due to the primordial de
Sitter bounce, and the strength of the effect mainly depends on
the duration of the de Sitter bounce after during the initial
expanding phase of the de Sitter bounce. The pre-inflationary
bounce is followed by a slow-roll quasi-de Sitter phase described
by vacuum $R^2$ gravity, which may or may not be followed by an
$f(R)$ gravity controlled reheating era. In all the cases, the
late-time era can be described in a viable way by some appropriate
$f(R)$ gravity. We will compute all the $f(R)$ gravities which can
realize the different patches of the Universe's evolution and
accordingly, we shall directly determine the energy spectrum of
the primordial gravitational waves for the resulting theories.

This paper is organized as follows: In section II we present in
brief our proposal for the primordial era of our Universe. We
describe in detail the three evolution patches of our Universe
primordially, which consist of a pre-inflationary de Sitter
bounce, followed by a quasi-de Sitter era, followed by a
geometrically generated reheating era. In section III, we discuss
how the different patches of our Universe's evolution can be
generated by $f(R)$ gravity, and we also discuss the qualitative
effects of the geometrically realized reheating era on the
inflationary era. In section IV we study several theoretical
scenarios and their predictions for the energy spectrum of the
primordial gravitational waves. The conclusions follow in the end
of the article.

\section{Pre-inflationary de Sitter Bounce and Primordial Evolution}

Let us discuss the scenario we propose in this work, which is
based on the fact that the Universe pre-inflationary was
experiencing a de Sitter bounce, which is followed by a quasi-de
Sitter era. Accordingly, after the quasi-de Sitter era we will
assume that the Universe enters a reheating era with constant
equation of state (EoS) parameter $w$, followed by the standard
patches of evolution, namely a canonical reheating era with EoS
parameter $w=1/3$ and finally the matter and dark energy era.
Giving the Universe's evolution in distinct patches is the best we
can do as cosmologists, since it is not possible to find the exact
scale factor which describes the Universe, this extends beyond the
reach of the human mind. Hence, assuming several evolutionary
patches for the Universe is the best that we can do, and in fact,
some of these patches may be directly determined, as it happens
with the dark energy era and also may happen with the inflationary
and post-inflationary era, via the future stage 4 CMB experiments
and the future gravitational waves experiments.
\begin{figure}[h!]
\centering
\includegraphics[width=20pc]{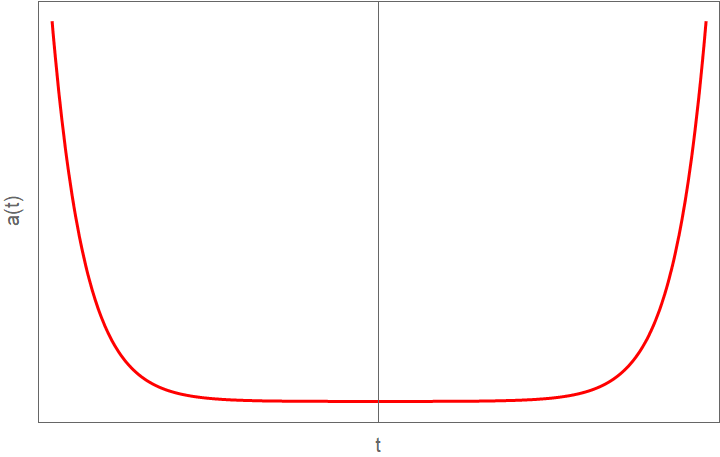}
\includegraphics[width=20pc]{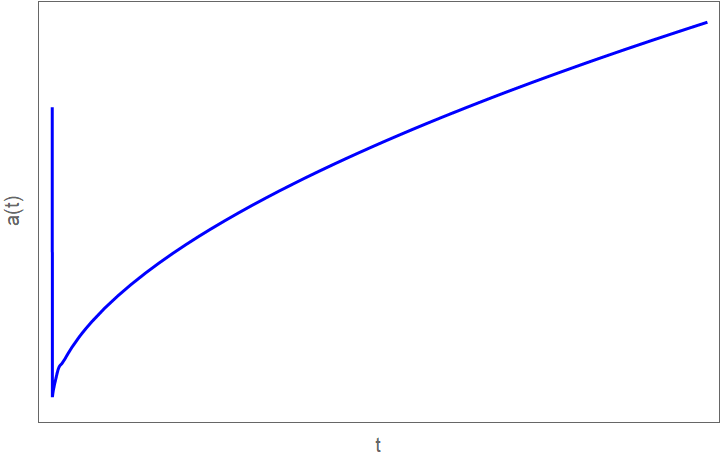}
\includegraphics[width=20pc]{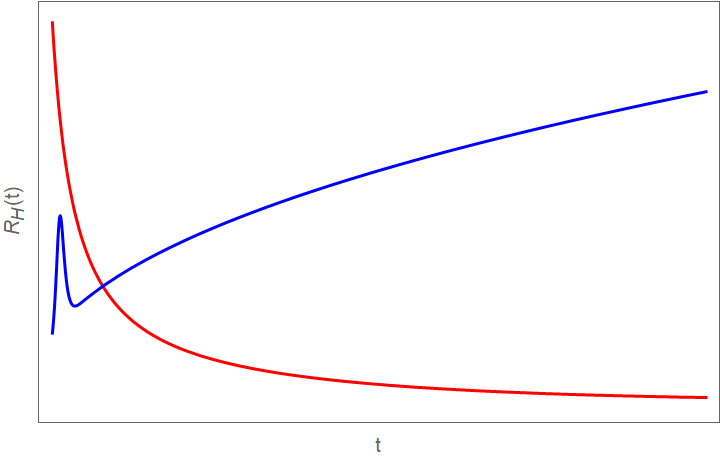}
\caption{Upper plots: The scale factor (\ref{unifiedscalefactor})
(blue curve) and the de Sitter bounce scale factor (red curve) vs
the cosmic time. Bottom plot: the Hubble radius
$R_H=\frac{1}{a(t)H(t)}$ as a function of the cosmic time for the
scale factor (\ref{unifiedscalefactor}) (blue curve) and for the
de Sitter bounce (red curve).} \label{plot1}
\end{figure}
However, pre-inflationary evolution patches are quite hard to be
probed, and our proposal in this paper is that these
pre-inflationary eras may have a direct effect on the energy
spectrum of the primordial gravitational waves, causing a
significant damping of the spectrum. In this line of research, let
us quote the scale factor for the pre-inflationary, inflationary
and the first moments of the post-inflationary epoch, which is,
\begin{equation}\label{unifiedscalefactor}
a(t)=a_b\cosh (j t)e^{-t/t_i}+a_ie^{H_0 t-H_i t^2}+a_w
\frac{t}{t_0}^{\frac{2}{3 (w+1)}}\, ,
\end{equation}
and let us explain the different patches for the above evolution.
The first term describes the de Sitter bounce
\cite{Matsui:2019ygj}, which is followed by the quasi-de Sitter
inflationary epoch described by the second term, followed by the
constant EoS parameter $w$ reheating epoch, described by the third
term. Now, $a_b$, $a_i$ and $a_w$ denote the size of the Universe
at the beginning of the de Sitter bounce, at the beginning of the
inflationary era and at the beginning of the reheating era with
constant EoS parameter $w$. The parameters $j$, $H_0$ and $H_i$
have mass dimensions eV, eV and eV$^2$, while the time instances
$t_i$ and $t_0$ are characteristic and denote the time that the
bounce ends and the time instance that the inflationary era ends.
So for cosmic times $t\ll t_i$, the exponential term is
practically equal to unity, while for times $t\geq t_i$ the
exponential term causes a damping of the first term, thus the
other two start to dominate. In the left and right upper plots of
Fig. \ref{plot1} we plot the scale factor
(\ref{unifiedscalefactor}) vs the cosmic time (blue curve) and the
de Sitter bounce scale factor described by the first term (red
curve). Also in the bottom plot we present the Hubble radius
$R_H=\frac{1}{a(t)H(t)}$ as a function of the cosmic time for the
scale factor (\ref{unifiedscalefactor}) (blue curve) and for the
de Sitter bounce (red curve). As it can be seen, the scale factor
(\ref{unifiedscalefactor}) is described by a bounce
pre-inflationary, in which case the scale factor decreases, and
the Hubble radius increases, after that the Universe experiences a
short period of acceleration, in which case the Hubble radius
decreases, and this short acceleration period is followed by a
deceleration period, in which case the Hubble radius starts to
decrease again. The blue curve has exactly the behavior described
by the scale factor (\ref{unifiedscalefactor}), so the Universe
starts with a per-inflationary bounce, followed by a short period
of inflation, followed by a a power-law evolution with constant
EoS parameter. For the plots we assumed that $w=0$ and this is
also what we will assume for the rest of the article. Hence
basically, the reheating era is abnormal and has an EoS parameter
$w=0$, different from $w=1/3$ which describes an ordinary
reheating era.

\section{Inflation and Post-inflation Evolution with $f(R)$ Gravity}

Let us now proceed in the modified gravity description of the
cosmological evolution we presented in the previous section. Our
basic assumption is that $f(R)$ gravity controls the whole
evolution, from the pre-inflationary era to the late-time era. It
is $f(R)$ gravity which realizes the various evolutionary patches
we described in the previous section. Schematically, the $f(R)$
gravity which realizes the evolution patches which we described in
the previous section will have the following form,
\[
f(R)=\left\{
\begin{array}{ccc}
F_B(R)& R\geq R_B\, ,  \\
  R+\frac{R^2}{6M^2}& R\sim R_I\, ,  \\
  F_w(R) & R\sim R_{PI}\ll R_I\, ,   \\
  F_{DE}(R) & R\sim R_0\ll R_{PI}\, , \\
\end{array}\right.
\]
with $R_I$ stands for the curvature scale of inflation, at the
first horizon crossing, $R_{PI}$ stands for the post-inflationary
curvature scale during the abnormal reheating era, and $R_B$ is
the curvature scale near the bouncing point, when the Universe is
described by the de Sitter bounce. Accordingly, $F_B(R)$ is the
$f(R)$ gravity which realizes the de Sitter bounce, $F_w(R)$ is
the $f(R)$ gravity which realizes the abnormal reheating era and
$F_{DE}(R)$ is the $f(R)$ gravity which realizes the late-time
era. Note that for the bouncing and inflationary eras, the matter
content of the Universe plays no fundamental role, so these are
vacuum $f(R)$ gravities, however the matter content
synergistically with $f(R)$ gravity realize the abnormal reheating
and late-time evolution of the Universe. Note that the quasi-de
Sitter inflationary evolution is realized by an $R^2$ gravity, see
for example Ref. \cite{Odintsov:2021wjz} for further details on
this issue. Now let us use some well known reconstruction
techniques in order to find the forms of $F_w(R)$ and $F_B(R)$.
The first was calculated in detail in Ref.
\cite{Oikonomou:2022yle}, in the presence of matter and radiation
fluids, and the result was found to be,
\begin{equation}
\label{newsolutionsnoneulerssss} F_{w}(R)=\left
[\frac{c_2\rho_1}{\rho_2}-\frac{c_1\rho_1}{\rho_2(\rho_2-\rho_1+1)}\right]R^{\rho_2+1}
+\sum_i
\left[\frac{c_1S_i}{\rho_2(\delta_i+2+\rho_2-\rho_1)}\right]
R^{\delta_i+2+\rho_2}-\sum_iB_ic_2R^{\delta_i+\rho_2}+c_1R^{\rho_1}+c_2R^{\rho_2}\,
,
\end{equation}
where $c_1,c_2$ are integration constants, and also $\delta_i$ and
$B_i$ are,
\begin{equation}
\label{paramefgdd}
\delta_i=\frac{3(1+w_i)-23(1+w)}{3(1+w)}-\rho_2+2,\,\,\,B_i=\frac{S_i}{\rho_2\delta_i}
\, ,
\end{equation}
with $i=(r,m)$, while $a_1$ and $a_2$, $S_i$ and $A$ are equal to,
\begin{equation}
\label{apara1a2} a_1=\frac{3(1+w)}{4-3(1+w)},\,\,\,
a_2=\frac{2-3(1+w)}{2(4-3(1+w))},\,\,\,S_i=\frac{\kappa^2\rho_{i0}a_0^{-3(1+w_i)}}{[3A(4-3(1+w))]^{\frac{3(1+w_i)}{3(1+w)}}},\,\,\,A=\frac{4}{3(w+1)}\,
.
\end{equation}
Now let us focus on the calculation of $F_B(R)$, and we shall use
a well known reconstruction technique \cite{Nojiri:2009kx} in
order to find this. The $f(R)$ gravity action in vacuum is,
\begin{equation}\label{action1dse}
\mathcal{S}=\frac{1}{2\kappa^2}\int
\mathrm{d}^4x\sqrt{-g}\,f(R)+\mathcal{S}_m,
\end{equation}
with $\kappa^2$ being $\kappa^2=8\pi G=\frac{1}{M_p^2}$, and $G$
is Newton's constant, with $M_p$ denoting the reduced Planck mass.
In the metric formalism, the field equations are,
\begin{equation}\label{eqnmotion}
f_R(R)R_{\mu \nu}(g)-\frac{1}{2}f(R)g_{\mu
\nu}-\nabla_{\mu}\nabla_{\nu}f_R(R)+g_{\mu \nu}\square f_R(R)=0\,
,
\end{equation}
with $f_R=\frac{\mathrm{d}f}{\mathrm{d}R}$. For a flat
Friedmann-Robertson-Walker (FRW) spacetime, the Friedmann equation
reads,
\begin{equation}\label{frwf1}
-18\left (4H(t)^2\dot{H}(t)+H(t)\ddot{H}(t)\right)f_{RR}(R)+3
\left(H^2(t)+\dot{H}(t) \right)f_{R}-\frac{f(R)}{2}=0\, ,
\end{equation}
We are interested in realizing the de Sitter bounce patch of the
scale factor (\ref{unifiedscalefactor}), so basically,
\begin{equation}\label{scalefactorquasidesitter}
a(t)\simeq a_b\cosh (j t)\, ,
\end{equation}
near the bouncing point. Using the $e$-foldings number,
\begin{equation}\label{efoldpoar}
e^{-N}=\frac{a_b}{a}\, ,
\end{equation}
as a dynamical variable in our cosmological system, the Friedmann
equation takes the following form,
\begin{equation}
\label{newfrw1} -18\left [ 4H^3(N)H'(N)+H^2(N)(H')^2+H^3(N)H''(N)
\right ]f_{RR}(R)+3\left [H^2(N)+H(N)H'(N)
\right]f_R(R)-\frac{f(R)}{2}=0\, .
\end{equation}
Introducing the function, $G(N)=H^2(N)$, the Ricci scalar is
written as,
\begin{equation}\label{riccinrelat}
R=3G'(N)+12G(N)\, .
\end{equation}
and finally the Friedmann equation takes the final form,
 \begin{equation}
\label{newfrw1modfrom} -9G(N(R))\left[ 4G'(N(R))+G''(N(R))
\right]f_{RR}(R) +\left[3G(N)+\frac{3}{2}G'(N(R))
\right]f_R(R)-\frac{f(R)}{2}=0\, ,
\end{equation}
with $G'(N)=\mathrm{d}G(N)/\mathrm{d}N$ and
$G''(N)=\mathrm{d}^2G(N)/\mathrm{d}N^2$ and
$f_{RR}=\frac{d^2f}{dR^2}$. Thus the $F_B(R)$ gravity which
realizes the de Sitter bounce, can be found by solving Eq.
(\ref{newfrw1modfrom}). In our case, $G(N)$ has the following
form,
\begin{equation}\label{gnfunction}
G(N)=\left(\frac{j \sqrt{\frac{\exp (N)-b}{b+\exp (N)}} (b+\exp
(N))}{\exp (N)}\right)^2\, ,
\end{equation}
hence the Friedmann equation (\ref{newfrw1modfrom}), takes the
final form,
\begin{align}
\label{bigdiffgeneral1} \left(-18 j^2 R+72 j^4+R^2\right)
f_{RR}(R)+3 j^2 f_R(R)-\frac{f(R)}{2}\, ,
\end{align}
which can be solved analytically, and the resulting function
$F_B(R)$ which realizes the de Sitter bouncing era has the
following form,
\begin{equation}\label{desitterbouncefr}
F_B(R)=c_1 \left(\sqrt{6 j^2-R} \sqrt{12 j^2-R}-9
j^2+R\right)^{\frac{\sqrt{3}}{2}} \sqrt{3 \sqrt{6 j^2-R} \sqrt{12
j^2-R}+15 \sqrt{3} j^2-2 \sqrt{3} R}\, ,
\end{equation}
with $c_1$ being an irrelevant integration constant. Finally, the
late-time era $f(R)$ gravity will be assumed to have the form,
\begin{equation}\label{starobinsky}
F_{DE}(R)=-\gamma \Lambda
\Big{(}\frac{R}{3m_s^2}\Big{)}^{\delta}\, ,
\end{equation}
where $m_s$ is $m_s^2=\frac{\kappa^2\rho_m^{(0)}}{3}$,
$\rho_m^{(0)}$ is the present day energy density of cold dark
matter, $\delta $ takes values $0<\delta <1$, and $\gamma$ is an
arbitrary dimensionless parameter, and finally $\Lambda$ is the
cosmological constant at present day. The late-time phenomenology
of the model (\ref{starobinsky}) in the presence of an $R^2$ and
subleading power-law terms was studied in Ref.
\cite{Odintsov:2021kup} and it is proven to be viable, so we will
not further study it here.

Having the functional forms of the total $f(R)$ gravity which
realize the various evolutionary patches of the Universe, in the
next section we shall investigate the effects of the bounce era on
the energy spectrum of the primordial gravitational waves. For the
calculation of the energy spectrum of the primordial gravitational
waves, we shall use the inflationary indices of the $R^2$ model,
which are,
\begin{equation}\label{tensotoscalarratio}
r=48\epsilon_1^2\, ,
\end{equation}
and the corresponding tensor spectral index for the $R^2$ gravity
is \cite{reviews1,Odintsov:2021kup},
\begin{equation}\label{tensorspectralindexr2ini}
n_T\simeq -2\epsilon_1^2\, ,
\end{equation}
with $\epsilon_1$ being the first slow-roll index
$\epsilon_1=-\dot{H}/H^2$. For the $R^2$ gravity, the first
slow-roll index is $\epsilon_1\simeq \frac{1}{2N}$, therefore we
have,
\begin{equation}\label{r2modeltensorspectralindexfinal}
n_T\simeq -\frac{1}{2N^2}\, ,
\end{equation}
and
\begin{equation}\label{tensortoscalarfinal}
r=\frac{12}{N^2}\, .
\end{equation}
Also the post-inflationary abnormal reheating era affects the
duration of the inflationary era, in the following way
\cite{Adshead:2010mc},
\begin{equation}\label{efoldingsmainrelation}
N=56.12-\ln \left( \frac{k}{k_*}\right)+\frac{1}{3(1+w)}\ln \left(
\frac{2}{3}\right)+\ln \left(
\frac{\rho_k^{1/4}}{\rho_{end}^{1/4}}\right)+\frac{1-3w}{3(1+w)}\ln
\left( \frac{\rho_{reh}^{1/4}}{\rho_{end}^{1/4}}\right)+\ln \left(
\frac{\rho_k^{1/4}}{10^{16}\mathrm{GeV}}\right)\, ,
\end{equation}
with $\rho_{end}$ and $\rho_{reh}$ being the energy density of the
Universe at the end of the inflationary era and at the end of the
reheating era respectively, $\rho_k$ is the energy density of the
Universe at the first horizon crossing, and $k_*$ is the pivot
scale $k_*=0.05$Mpc$^{-1}$. Hence, for the abnormal reheating
scenario with $w=0$ post-inflationary, the inflationary era is
either prolonged beyond 60 $e$-foldings, or it lasts for less
amount of time, depending on the reheating temperature. We shall
consider three reheating temperatures, $T_R=10^{12}$GeV,
$T_R=10^{7}$GeV and $T_R=10^{2}$GeV, and in Table \ref{table2} we
quote the duration of the inflationary era in terms of the
$e$-foldings number, and the values of the inflationary indices
for the three distinct reheating temperatures.
\begin{table}[h!]
  \begin{center}
    \caption{\emph{\textbf{Duration of Inflation and Inflationary Indices for Three Reheating Temperatures}}}
    \label{table2}
    \begin{tabular}{|r|r|r|r|}
     \hline
      \textbf{$e$-foldings and Inflationary Indices} & \textbf{$T_R=10^{12}$GeV} & \textbf{$T_R=10^{7}$GeV} & \textbf{$T_R=10^{2}$GeV}\\
           \hline
           $e$-foldings number $N$ & 65.3439 & 61.5063& 57.6687\\ \hline
 Tensor Spectral Index $n_T$ & -0.0000585503 & -0.0000660847 & -0.0000751727 \\ \hline
      Tensor-to-Scalar Ratio $r$ & 0.00281042 & 0.00317206 & 0.00360829\\ \hline
    \end{tabular}
  \end{center}
\end{table}
We shall use the results of Table \ref{table2} in the next section
for the calculation of the energy spectrum of the primordial
gravitational waves. We need to note that the transition from the
bounce epoch to the inflationary epoch is smooth and continuous as
it can be seen by looking the scale factor in Eq.
(\ref{unifiedscalefactor}). The realization of the scale factor
(\ref{unifiedscalefactor}) could be quite difficult to do in
standard single scalar field theory in the Einstein frame, this is
why we chose to realize this in the context of Jordan frame $f(R)$
gravity. In principle in Einstein frame scalar field theory, the
realization of the scale factor (\ref{unifiedscalefactor}) would
require complicated potentials and unnatural perfect fluids, with
the issue of viability of the inflationary scenario being apparent
and uncertain. This is why we chose the Jordan frame description,
in which both the viability of the inflationary era is guaranteed
and also the resulting $f(R)$ gravity model which realizes the
scale factor (\ref{unifiedscalefactor}) is quite elegant and
simple.

\section{The Energy Spectrum of the Primordial Gravitational Wave: Effects of the de Sitter Bounce}

In this section we shall quantitatively study the effect of the de
Sitter bounce pre-inflationary era on the energy spectrum of the
primordial gravitational waves. To this end we have to specify the
duration of the pre-inflationary bounce era. Since the Planck era
corresponds to a temperature of $10^{19}$GeV, and inflation is
believe to commence at $T\sim 10^{16}$GeV, we shall assume that
the pre-inflationary bounce epoch starts at $T\sim 10^{19}$GeV and
ends at $T\sim 10^{16}$GeV. We must translate this temperature
interval into a redshift interval, since this is essential for the
calculation of the damping factor, so using the relation
$T=T_0(1+z)$ \cite{Garcia-Bellido:1999qrp}, where $T_0$ denotes
the present day temperature $T_0=2.58651\times 10^{-4}$eV, the
temperature interval $T=[10^{16}-10^{19}]$GeV corresponds to the
redshift interval $z=[3.86621\times 10^{28},3.86621\times
10^{31}]$. After the end of the bouncing era, the inflationary era
commences which is assumed to occur at a temperature $T\sim
10^{16}$GeV and ends for example at a temperature $T_{end}\sim
10^{13}$GeV. The abnormal reheating era then commences which we
shall assume that it lasts until the temperature drops to the
order $T_{pr}=10^{12}$GeV. Translated in redshifts, the abnormal
reheating era lasts for the redshift interval $z=[3.86621\times
10^{15},3.86621\times 10^{16}]$. These redshifts intervals are
quite important for the calculation of the damping factor caused
by the $f(R)$ gravity during the pre-inflationary bounce and
during the abnormal reheating era.
\begin{figure}[h!]
\centering
\includegraphics[width=30pc]{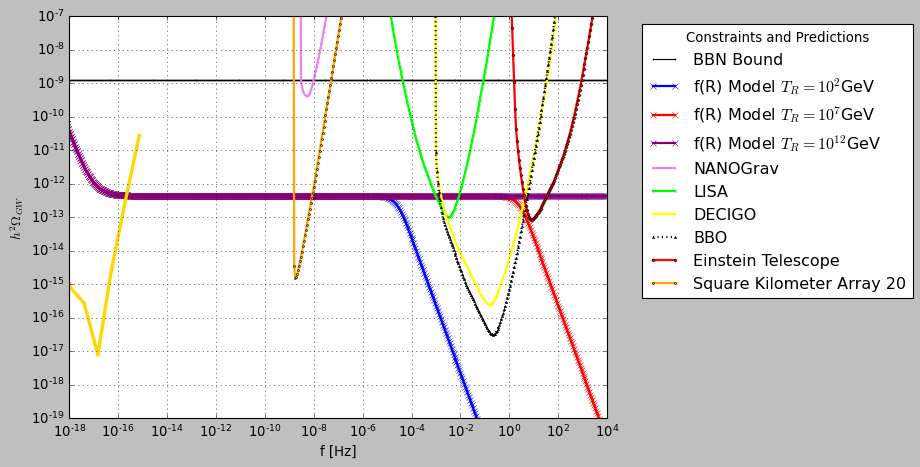}
\includegraphics[width=30pc]{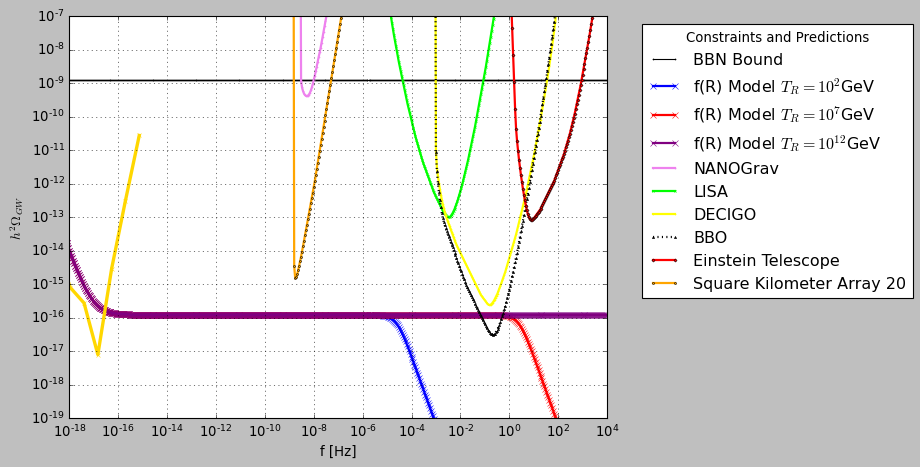}
\caption{The $h^2$-scaled gravitational wave energy spectrum for
the combined $f(R)$ gravity model with a primordial de Sitter
bounce era (upper plot) and without the pre-inflationary bounce
era (bottom plot), for three reheating temperatures, namely for
$T_R=10^{12}$GeV (purple curves), to $T_R=10^{7}$GeV (red curves)
and (blue curves) $T_R=10^{2}$GeV.} \label{plot2}
\end{figure}

At this point, let us review the method of extracting the overall
damping or amplification effect of the modified gravity on the GR
waveform. For details we refer to the review
\cite{Odintsov:2022cbm} and to the original paper
\cite{Nishizawa:2017nef} where the method appeared for the first
time. The crucial parameter that quantifies the overall effect of
modified gravity is the parameter $a_M$ which for $f(R)$ gravity
is defined,
\begin{equation}\label{amfrgravity}
a_M=\frac{f_{RR}\dot{R}}{f_RH}\, ,
\end{equation}
and the waveform of modified gravity in terms of the GR waveform
has the following form \cite{Nishizawa:2017nef,Arai:2017hxj},
\begin{equation}\label{mainsolutionwkb}
h=e^{-\mathcal{D}}h_{GR}\, ,
\end{equation}
where $h_{GR}$ denotes the GR waveform in which case $a_M=0$, and
$\mathcal{D}$ is equal to,
\begin{equation}\label{dform}
\mathcal{D}=\frac{1}{2}\int^{\tau}a_M\mathcal{H}{\rm
d}\tau_1=\frac{1}{2}\int_0^z\frac{a_M}{1+z'}{\rm d z'}\, .
\end{equation}
The details on the WKB method used to derive the above solutions
can be found in the review \cite{Odintsov:2022cbm}. Thus, the
energy spectrum for the $f(R)$ gravity is
\cite{Boyle:2005se,Nishizawa:2017nef,Arai:2017hxj,Nunes:2018zot,Liu:2015psa,Zhao:2013bba,Odintsov:2021kup,Odintsov:2022cbm},
\begin{align}
\label{GWspecfR}
    &\Omega_{\rm gw}(f)=e^{-2\mathcal{D}}\times \frac{k^2}{12H_0^2}r\mathcal{P}_{\zeta}(k_{ref})\left(\frac{k}{k_{ref}}
\right)^{n_T} \left ( \frac{\Omega_m}{\Omega_\Lambda} \right )^2
    \left ( \frac{g_*(T_{\rm in})}{g_{*0}} \right )
    \left ( \frac{g_{*s0}}{g_{*s}(T_{\rm in})} \right )^{4/3} \nonumber  \left (\overline{ \frac{3j_1(k\tau_0)}{k\tau_0} } \right )^2
    T_1^2\left ( x_{\rm eq} \right )
    T_2^2\left ( x_R \right )\, ,
\end{align}
with $k_{ref}=0.002$$\,$Mpc$^{-1}$ denoting the CMB pivot scale,
$n_T$ stands for the tensor spectral index and $r$ is as usual the
tensor-to-scalar ratio. Our main task in this section is to
numerically evaluate the parameter $\mathcal{D}$ for all the
redshifts up to the Planck era with redshift $z_p=3.86621\times
10^{31}$. The main contributions which are not trivial are
contributed by the redshift intervals $z=[3.86621\times
10^{15},3.86621\times 10^{16}]$ and $z=[3.86621\times
10^{28},3.86621\times 10^{31}]$ which correspond to the abnormal
reheating and the de Sitter bounce pre-inflationary era. The rest
of the redshift intervals contribute terms of the order of unity,
so we shall not discuss them here, we refer the reader to Ref.
\cite{Odintsov:2021kup} for details. Now recall that the de Sitter
bounce and the abnormal reheating era are generated by different
$f(R)$ gravities, which plays an important role for the
calculation, since the parameter $a_M$ is essentially different
for these two eras. Our numerical analysis indicates the
following, the parameter $\mathcal{D}$ for the redshift interval
$z=[3.86621\times 10^{15},3.86621\times 10^{16}]$ is
$\mathcal{D}=-14.5063$ while for the redshift interval
$z=[3.86621\times 10^{28},3.86621\times 10^{31}]$ it is equal to
$\mathcal{D}=8.169$. Thus the abnormal reheating leads to an
amplification of the GR waveform of the order
$\mathcal{O}(10^{6})$, however the de Sitter pre-inflationary
bounce causes a damping of the order $\mathcal{O}(10^{-4})$. Thus
the damping effect of the de Sitter pre-inflationary bounce is
significant. To have a quantitative idea on how the
pre-inflationary bounce affects the energy spectrum of the
primordial gravitational waves, in Fig. \ref{plot2} we plot the
$h^2$-scaled gravitational wave energy spectrum for the combined
$f(R)$ gravity model with a primordial de Sitter bounce era (upper
plot) and without the pre-inflationary bounce era (bottom plot),
for three reheating temperatures, namely for $T_R=10^{12}$GeV
(purple curves), to $T_R=10^{7}$GeV (red curves) and (blue curves)
$T_R=10^{2}$GeV. As it can be seen in the two plots of Fig.
\ref{plot2}, in the absence of the pre-inflationary bounce, the
gravitational wave signal is detectable from all the gravitational
wave experiments for all the reheating temperatures, however the
effect of the pre-inflationary bounce causes significant damping,
and in effect it will be detectable only from one detector, the
BBO, and only if the reheating temperature is larger than
$10^{7}$GeV. This scenario is of profound importance, although it
is rather model dependent. It is important because it teaches us
that a plethora of future scenarios might indicate several things
for the primordial era. In our case, if a primordial bounce took
place, the signal will be detected only by one detector, and only
if the reheating temperature is sufficiently large. If the
reheating temperature is low, then no signal will be detected. On
the contrary if the pre-inflationary bounce is absent, then the
signal of stochastic gravitational waves will be detected by all
the detectors, or by some detectors. Thus the detection or non
detection of the signal may yield information on the underlying
scenario, and specifically, the theory that controls inflation and
the reheating era, and also whether an exotic pre-inflationary
epoch took place. The latter will be more apparent to have
occurred, if the signal is detected by some but not all the
detectors, so in a specific frequency range. So overall, the plot
thickens with primordial gravitational waves. We need to note that
for all the cases we studied, the low-frequency predictions should
be disregarded because our analysis is valid for modes that became
subhorizon shortly after the inflationary era, so for frequencies
larger than $f>10^{-10}\,$Hz.

\section{Conclusions}

In this paper we calculated the effect of a primordial de Sitter
bounce on the energy spectrum of the primordial gravitational
waves. Particularly we assumed that pre-inflationary the Universe
experienced a de Sitter bounce phase, which is followed by a
quasi-de Sitter slow-roll inflationary era, followed by a constant
EoS parameter reheating era, which is followed by the standard
radiation and matter domination eras, up to the late-time
acceleration eras. So we assumed several cosmological evolution
patches for the Universe, and we also assumed that these evolution
patches are realized by vacuum $f(R)$ gravity or synergistically
by $f(R)$ gravity and the perfect matter fluids. Specifically,
vacuum $f(R)$ gravity generates the primordial bounce and the
slow-roll inflationary eras, while the synergy of $f(R)$ gravity
and the perfect fluids, realize the abnormal reheating era and the
late-time acceleration eras. The quasi-de Sitter inflationary era
is realized by an $R^2$ while for the rest of the eras, we
investigated which $f(R)$ gravity may realize these eras, by using
well-known reconstruction techniques. Then by having available the
$f(R)$ gravities which realize each era, we calculated the overall
effect of the primordial de Sitter bounce on the spectrum of the
primordial gravitational waves, and we compared the results with
the case in which the primordial bounce is absent. As we
demonstrated, the primordial de Sitter bounce causes a severe
damping of the energy spectrum of the primordial gravitational
waves, which in the model we studied the signal would be
detectable only by one future gravitational wave detector. This
result is of profound importance, for the following reason: if a
signal is detected in some but not all the future gravitational
wave detectors, this scenario may be explained by the presence of
a primordial bounce cosmology. In fact, this maybe more plausible
than other scenarios which may explain the absence of a signal in
some frequency range. This is because if the signal is detected
only by some detectors in some frequencies, this signifies
probably a global mechanism affecting all frequencies, and not a
physical mechanism corresponding to a specific frequency range,
such as supersymmetry breaking at some time instance during the
reheating era. Thus, with this paper we offered another
perspective on primordial gravitational waves physics, and one
thing is certain, the detection of a signal may have multiple
theories and effects that may describe it. Therefore, the plot
thickens with primordial gravitational waves. Before closing, let
us further comment on an interesting issue.  It is known that in
the bouncing cosmology scenario, there will be a huge increase of
anisotropies, which eventually could give rise to large secondary
gravitational waves sources, which are induced by linear scalar
perturbations. This may even exceed the linear gravitational
waves. This effect could also be investigated in future extensions
of this work and requires a combination of linear and non-linear
sources of gravitational waves. In this article however we merely
focused on linear gravitational waves effects, so for modes with
wavelengths well above 10$\,$Mpc.

\end{document}